\documentclass[%
 amsmath,amssymb,amsfonts,
 aps,prl,twocolumn
]{revtex4-1}

\usepackage{amsmath}
\usepackage[dvips]{graphicx}% Include figure files
\usepackage{wrapfig}
\usepackage{bm}% bold math
\usepackage[percent]{overpic}

\begin{document}

%\preprint{APS/123-QED}

\title{{\it Ab initio} approach to the lattice softening of an Al slab driven by collective electronic excitations after ultrashort laser pulse irradiation}% Force line breaks with \\

\author{Hiroki Katow}
\author{Yoshiyuki Miyamoto}%
\affiliation{%
 Research Center for Computational Design of Advanced Functional Materials, National Institute
of Advanced Industrial Science and Technology (AIST), Central 2, Tsukuba, Ibaraki 305-8568, Japan
}%


%\affiliation{
%\affiliation{
 %Third institution, the second for Charlie Author
%\affiliation{%


\date{\today}% It is always \today, today,

\begin{abstract} 
Recent advances in ultrashort laser pulse techniques have opened up a wide variety of applications in both fundamental physics and industrial fields. 
In this work, $ab$ $initio$ molecular dynamics simulations based on time-dependent density functional theory revealed a steady deceleration of lattice distortion propagation in an aluminum slab with increasing laser pulse intensity.
Analysis of the interatomic force revealed a significant reduction in the harmonic terms and non-monotonic growth of anharmonicity.
This behavior was characterized by spatially non-uniform force screening by plasmons, which is missing from Born--Oppenheimer molecular dynamics,
and is consistent with the current interpretation of laser-induced periodic structure patterning.
This work provides a semi-quantitative criterion for modifying the phonon properties of non-equilibrium systems.  
%\begin{description}
%\end{description}
\end{abstract}

\pacs{Valid PACS appear here}% PACS, the Physics and Astronomy
\maketitle

%\tableofcontents

%\section{\label{sec:intro}introduction}
Material processing techniques using ultrafast intense laser pulses have been widely used in both fundamental physics and industrial fields \cite{SC14}.
 In comparison to processing methods based on nanosecond laser pulses, the use of sub-picosecond laser pulse irradiation reduces the thermal and/or energy diffusion into the surrounding medium, which leads to high energy efficiency and fine spatial resolution during patterning.
The realization of greater efficiency and finer resolution than the laser wavelength has been vigorously sought.
Although more than three decades have passed since early experimental reports \cite{SSB87,KS87}, elucidating the material properties under or after pulse irradiation remains at the cutting edge of condensed matter physics.
However, the extremely non-equilibrium and multiscale nature of ablation processes continues to hinder research in this area.

Reducing the interatomic potential and energy diffusion to the medium would be favorable for improving the spatial resolution and energy efficiency during laser patterning.
Various mechanisms of lattice property modulation have been proposed to investigate the formation of sub-wavelength structures during ablation processes.
It is widely known that such structures are formed and that their spatial periodicity depends on the laser pulse duration.
When nanosecond laser pulses are used, the periodicity is close to the incident laser wavelength. 
This is considered to originate from the interference between the incident and reflected laser light \cite{JKSU81,KB82}.
Meanwhile, femtosecond laser pulse irradiation generates grating structures whose periodicity is one order of magnitude smaller than the laser wavelength \cite{SHTNO09,YMK03,BH03,CHR03,RCHP02,MM06,SKQH03,WHJSM10}. 
Numerous mechanisms to explain these phenomena have been proposed, some of which have considered the contribution of plasmonic excitations \cite{RCHP02,BH03,BSRHTRC06,MM08,VMG07,BRK09}.
The direct measurement of lattice properties during ultrafast processes is difficult and fundamental quantities such as the interatomic force constants are only given for thermalized equilibrium system \cite{KLDLF08}. 
Consequently, further information regarding laser--matter interactions would be of great value.

Since ablation is a multiscale process, previous theoretical approaches have ranged from macro- to microscopic models and the descriptions of the systems have also varied. 
Hydrodynamic models with a nanosecond time scale and micrometer spatial scale have been reviewed previously \cite{SEP13}.
Classical molecular dynamics (MD) simulations have also been performed for sub-micrometer-scale structures of metals \cite{IKLRCES13, NSS12}.
Many studies using quantum mechanical approaches have assumed thermalization of the subsystems \cite{BSNGA16,WBEV16}. 
In a recent advance, a first-principles study highlighted the importance of electronic enthalpy in ablation processes based on finite-temperature density functional theory (DFT) \cite{TS18}. 

In this Letter, we demonstrate the application of {\it ab initio} Ehrenfest molecular dynamics (EMD) simulations based on time-dependent density functional theory (TDDFT) to investigate the laser-driven suppression of interatomic forces and the volume expansion of crystals. 
The investigated system was a thin slab comprising nine layers of Al atoms.  
The (111) surfaces of the Al fcc structure is exposed to the vacuum. 
We applied an ultrashort laser pulse of infrared light with a wavelength of 800 nm, a full width at half-maximum (FWHM) of 10 fs, and a field amplitude ranging from 0.0 to 3.0 $\mathrm{V/\AA}$. 
To analyze the force field, we employed a quasi-one-dimensional model in which atoms were coupled to their first nearest neighbors  via a potential expanded by the third order  of interatomic distance.
The force constants were fitted such that the model reproduced the EMD trajectory.
We confirmed the volume expansion of the slab and significant suppression of the harmonic terms, which amounted to 38\% of the initial values when averaged over the layers after irradiation. 
The corresponding anharmonic terms were also determined.
The force suppression was significant on the surface layers, in contrast to the case of Born--Oppenheimer MD (BOMD) simulations with finite electron temperature. 
Based on a phenomenological analysis, we attributed this discrepancy to force screening by plasmonic excitations, which is absent from the BOMD framework.

%Moreover, we find high frequency components in the Hellmann-Feynman force of atoms which we attributed to high harmonic oscillation (HHO), plasmonic excitation, and HHO-plasmon complexes.
%We estimate the magnitude of screening effect based on a model where ions and electronic density linearly couple.  
%We therefore conclude that the bizarre nature of EMD results is a consequence of screening effect to ionic motion by those excitations.  
%The analysis of electronic excitation enables us to understand the finite size effect and extrapolate our results to larger systems.
%According to our knowledge our approach is the first attempt to extract force constant modulation in such condition.

%\subsubsection{Introduction of theoretical framework}
We first introduce our theoretical approach to treating the ultrafast dynamics of the electronic and lattice systems.
The theoretical description of the real-time evolution of the electronic system was based on TDDFT.
In TDDFT, the electronic state at each time step is obtained by solving the time-dependent Kohn--Sham equation for one-particle orbitals \cite{RG84}. 
We used the local-density approximation (LDA) with a Perdew--Zunger-type exchange-correlation functional \cite{PerdewZunger}.
The fourth-order Suzuki--Trotter-type time evolution operator \cite{S92,SM99} was used to ensure numerical accuracy and unitarity of time evolution. 
Potentials between time steps were interpolated using the railway curve interpolation scheme for numerical accuracy and time reversibility \cite{SM99}.
  
 \begin{figure}%[htbp]
%\hspace{5mm}
\begin{tabular}{c}
	\begin{minipage}{1.0\hsize}
	\begin{center}
	\begin{overpic}[percent,width=6.5cm]{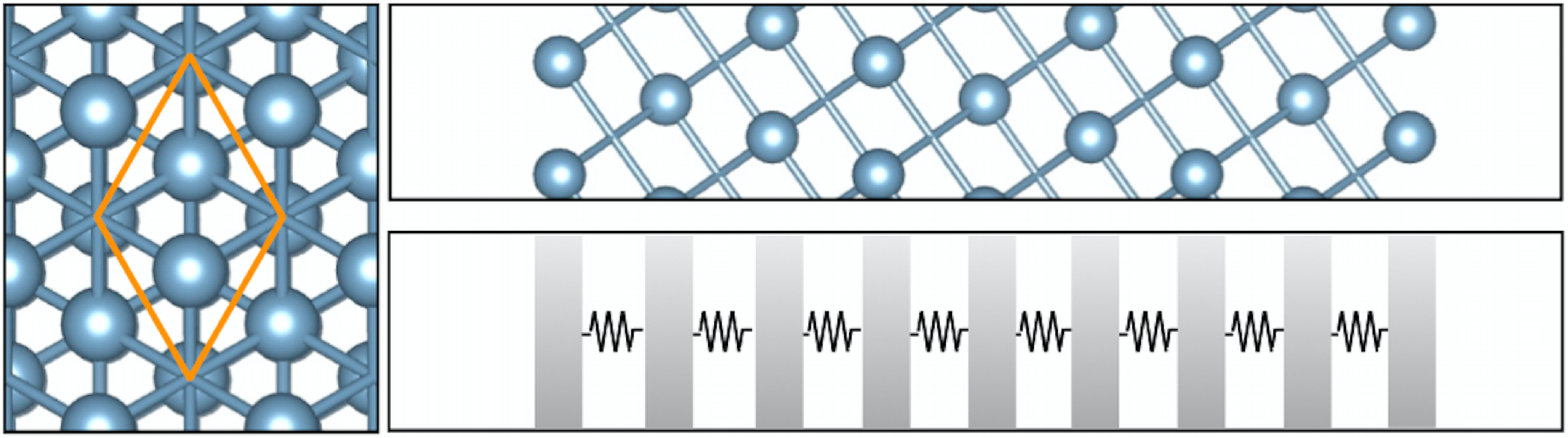}
	%\begin{overpic}[percent,grid,width=6.5cm]{./schematic.eps}
	%\put(0,24){\colorbox{white}{(a)}}
	\put(-6,24){(a)}
	\put(26,24){(b)}
	\put(26,9){(c)}
	\end{overpic}
	\end{center}
	\end{minipage}
	\\
	\begin{minipage}{1.0\hsize}
	\begin{center}
	%\begin{overpic}[percent,grid,width=5cm]{./propagate.eps}
	\begin{overpic}[percent,width=5cm]{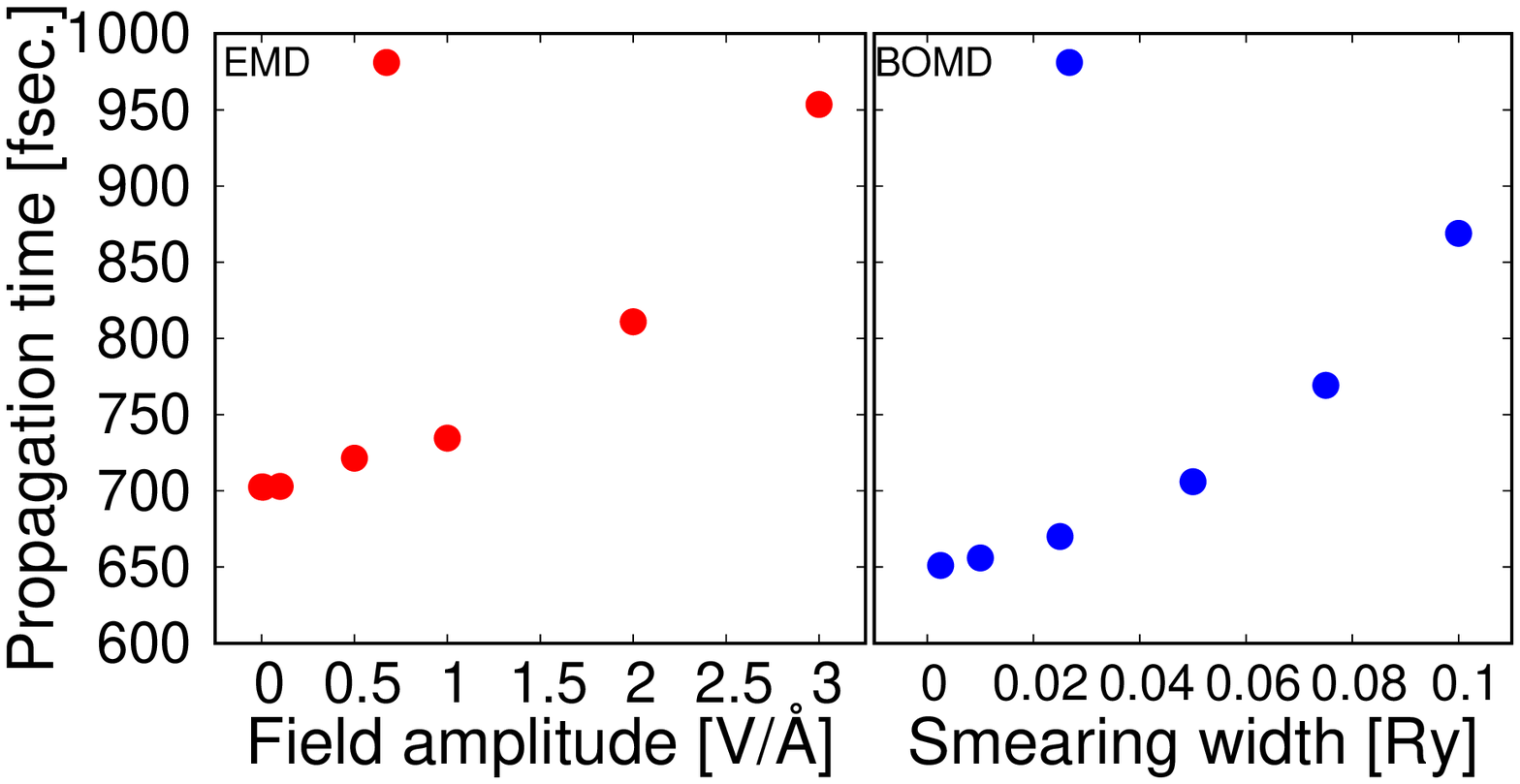}
		\put(-17,42){(d)}
	\end{overpic}
	\end{center}
	%\hspace{0.5cm}
	\end{minipage}
\end{tabular}
\caption{Crystal structure of the atomically thin aluminum slab model. (a) View of the Al(111) surface. The boundary of a unit cell is indicated by the orange line. (b) Cross section of the slab along the $z$ axis. (c) Schematic depiction of the quasi-one-dimensional model. Neighboring layers are bounded by an interatomic potential. (d) Propagation time of atomic displacement from one side of the slab to the other in EMD (left panel) and BOMD (right panel) simulations.} 
\label{fig:structure}
\end{figure}

In Fig.~\ref{fig:structure} we show the crystal structure of the thin aluminum slab used in the present simulation. 
We took the $xy$ plane as parallel to the slab. 
Fig.~\ref{fig:structure}(a) depicts the in-plane hexagonal unit-cell structure and Fig.~\ref{fig:structure}(b) shows a cross section of the slab along the $z$ axis.
The (111) surfaces of the fcc Al crystal were exposed to vacuum layers and the slab was composed of nine atomic layers.
The lengths of the $a$ ($b$) and $c$ axes were 5.303 bohr and 60.55 bohr, respectively.
The cell parameters were fixed in the EMD simulations.
Nine Al atoms were contained in the unit cell.
We used the Troullier--Martins-type  pseudopotential \cite{TM} and a 16$\times$16$\times$1 $k$-point mesh. 
The plane-wave vector cutoff was 35 Ry for the basis set and 562.5 Ry for the charge density distribution.
The time step for the time-dependent simulation was 3.63 attoseconds.
The electronic system was coupled to the external field by the length gauge $V = e\bm{r}\cdot \bm{E}$, where $e$ is the electronic charge, $\bm{r}$ is the electron position, and $\bm{E}$ is the external electric field with the polarization vector parallel to the $z$ axis. 
We applied an ultrashort laser pulse that can be analytically expressed as the product of a Gaussian and a sinusoidal function.
The FWHM was 10 fs and the frequency was 375 TH$_{\rm Z}$,  which corresponds to a wavelength of 800 nm. 
The maximum field amplitude ranged from 0.0 to 3.0 V/$\mathrm{\AA}$ in 0.5 V/$\mathrm \AA$ intervals.

\begin{figure}%[htbp]
\begin{tabular}{c}
\begin{minipage}{1.0\hsize}
	%\begin{overpic}[percent,grid,width=7cm]{./Layer.eps}
	\begin{overpic}[percent,width=7cm]{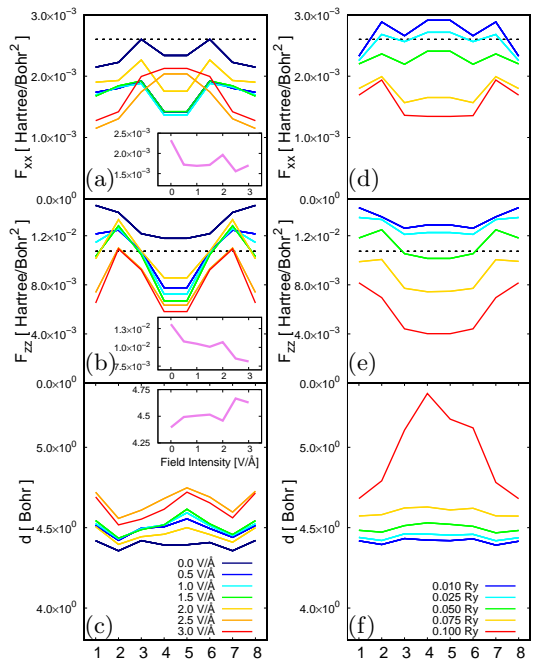}
	\put(11.7,73){(a)}
	\put(11.7,45){(b)}
	\put(11.7,5){(c)}
	\put(51.7,73){(d)}
	\put(51.7,45){(e)}
	\put(51.7,5){(f)}
	\end{overpic}
\end{minipage}
\end{tabular}
\vspace{0.5cm}
\caption{Second-order force constants $F_{xx}$ and $F_{zz}$ and interlayer distance $d$ for each layer for the (a)--(c) EMD and (d)--(f) BOMD simulations. 
The horizontal axes indicate the layer indices.
The dotted lines indicate the force constants of bulk fcc aluminum calculated using the relation between the experimental sound velocity $c_s$ and force constant $f$ for the one-dimensional model, $c_s=(fa^2/M)^{1/2}$, where $a$ is a lattice constant and $M$ is the mass of an Al atom.  The insets in (a), (b), and (c) show the average values over layers with respect to the laser field amplitude.
%Boltzmann const. = 1.380649 \times10^-23 J/K^-1
%Rydberg = 4.35974\times10^-18 J
}
\label{fig:force}
\end{figure}

We conducted simulations to measure the propagation time for in-plane atomic displacement, which was initially induced to an outermost atom.
This quantity was defined as the time required for the in-plane displacement of the opposite outermost atom to show its first peak. 
The left panel of Fig.~\ref{fig:structure}(d) shows the propagation time for various field amplitudes in the EMD simulations.
For comparison, we performed BOMD simulations with Fermi--Dirac smearing using the {\it Quantum ESPRESSO} package \cite{G09}.
We used the Troullier--Martins-type pseudopotential \cite{TM}, a 16$\times$16$\times$1$k$-point mesh, a plane-wave cutoff of 20 Ry, a charge density cutoff of 80 Ry, and a time step of 0.48 fs.
Provided that the smearing adequately approximate the effect of finite electronic temperature, we conducted our BOMD simulations with different smearing widths from 0.01 to 0.1 Ry, as shown in the right panel of Fig.~\ref{fig:structure}(d).
The discrepancy between the EMD and BOMD simulations at the lowest field amplitude and smearing width may be attributable to the different methods used to assign the occupations and parameter settings for the pseudopotentials, although it was outside the scope of this study to resolve this discrepancy by fine-tuning these parameters. 
The steady increase in the propagation time indicates the reduction of the interatomic potential. 
For quantitative analysis, we constructed a quasi-one-dimensional model in which each atomic layer  was coupled to its first nearest neighbor via an interatomic potential $U$ expanded by the third order  of interatomic distance:
\begin{eqnarray}
U(F^{(2)},F^{(3)},\bm{R}) = \sum_{m,n,i,j} F^{(2)}_{mn,ij}\xi_{mn,i}\xi_{mn,j}\nonumber \\
	+\sum_{m,n,i,j} F^{(3)}_{mn,ijk}\xi_{mn,i}\xi_{mn,j}\xi_{mn,k}.
\label{eq:potential}
\end{eqnarray}
Here, the interatomic distance $\xi_{mn,i} = u_{m,i}-R_{m,i}-u_{n,i}+R_{n,i} $,  where $u_{m,i}$ is the $i$-th component of the $m$-th atomic coordinate and $R_{m,i}$ is the corresponding equilibrium position.   
Fig.~\ref{fig:structure}(c) shows a schematic expression of this model. 
$F^{(2)}=\{F^{2}_{mn,ij}\}$, $F^{(3)}=\{F^{(3)}_{mn,ijk}\}$, and $\bm{R}=\{R_{m,i}\}$ are fitting parameters.
We projected the potential $U$ such that it satisfies hexagonal symmetry.
Thus, Eq.~(\ref{eq:potential}) is reduced to 
$U(F^{(2)},F^{(3)},\bm{R}) = \sum_{mn}\{F^{(2)}_{mn,xx}(\xi_{mn,x}^2+\xi_{mn,y}^2)+F^{(2)}_{zz}\xi_{mn,z}^2+F^{(3)}_{xxy}(\xi_{mn,x}^2\xi_{mn,y}-\xi_{mn,y}^3/3)+F^{(3)}_{mn,xxz}(\xi_{mn,x}^2+\xi_{mn,y}^2)\xi_{mn,z}+F^{(3)}_{zzz}\xi_{mn,z}^3\} $, which is characterized by five independent force constants. 
The potential $U$ is assumed to be invariant under inversion of the $z$ axis $\xi_{mn,z} \rightarrow -\xi_{mn,z}$.
In our fitting procedure, the evaluation function was defined as the square of the difference between the acceleration of atoms extracted from the EMD trajectory and those constructed by $F^{(2)}$, $F^{(3)}$, $\bm{R}$.
For the fitting, we ran the EMD simulation for 392 fs and randomly selected 50 snapshots of acceleration to construct the evaluation function.
All of the atomic positions were initially displaced from their equilibrium positions by 5\% of the lattice constant.
The displacement vector was set to be antiparallel to those of neighboring atoms.  
We omitted the high-frequency component of acceleration from the EMD trajectory by applying a 100  TH$_{\rm Z}$ cutoff prior to parameter optimization using the Fletcher--Reeves optimization method.
Additional details are provided in Sec.~S.I of the Supplemental Material \cite{Suppl}.

\begin{figure}%[tb]
\begin{tabular}{c}
\begin{minipage}{1.0\hsize}
%\begin{overpic}[percent,grid,width=7cm]{./F3rd.eps}
\begin{overpic}[percent,width=7cm]{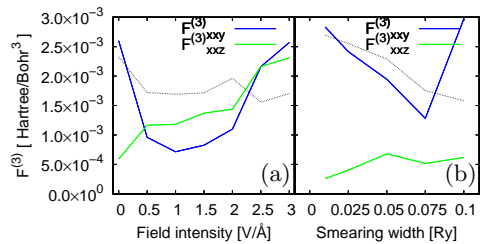}
\put(48,12){(a)}
\put(83,12){(b)}
\end{overpic}
\end{minipage}
\end{tabular}
\vspace{0.5cm}
\caption{Absolute values of the third-order force constants $F^{(3)}_{xxy}$ and $F^{(3)}_{xxz}$ averaged over layers for the (a) EMD and (b) BOMD simulations. For comparison, the values of $F^{(2)}_{xx}$ are indicated by black dotted lines.
}
\label{fig:force3rd}
\end{figure}

In Fig.~\ref{fig:force}(a), (b), and (c), we show the fitted second-order force constants and interlayer distances obtained from the EMD simulations. 
%We compare them with BOMD simulations with different Fermi-Dirac smearing width ranging from 0.01 Ry to 0.1 Ry as shown in Fig.~\ref{fig:force} (d), (e) and (f).
We compared the results with those from the BOMD simulations, as shown in Fig.~\ref{fig:force} (d), (e), and (f).
The total time for the BOMD simulation was 1.45 ps.
We observed significant suppression of $F^{(2)}$ in both cases.
The reductions in $F_{zz}^{(2)}$ averaged over the layers amounted to 38\% and 56\% of the initial values for the EMD and BOMD simulations, respectively.
Although there is no direct experimental report for observing these quantities, a neutron scattering experiment involving bulk aluminum revealed a 4.8\% reduction in phonon frequency and 10\% reduction in force constants for first nearest neighbors  when the temperature was increased from 10 to 775 K \cite{KLDLF08}.
It is counterintuitive that the decrease in $F^{(2)}_{xx}$ appears to saturate while the propagation time of atomic displacement exhibits a steady increase in Fig.~\ref{fig:structure}(d).
The absolute values of $F^{(3)}_{xxy}$ and $F^{(3)}_{xxz}$ averaged over the layers are plotted in Fig.~\ref{fig:force3rd}, revealing steady growth of these quantities with increasing field amplitude after the reduction in $F^{(2)}_{xx}$ became moderate.
We can deduce that the delay in propagation was partially due to the increase in $F^{(3)}$ at strong field intensity.
The values of $F^{(3)}_{zzz}$ did not converge under our optimization conditions and are therefore not shown. 
The spatial dependency of $F^{(3)}$ is summarized in Fig.~S.1 of Sec.~S.II of the Supplemental Material \cite{Suppl}.
Although the spatial non-uniformity of the force constants was large, i.e., the finite size effect was significant, our results provide a semi-quantitative criterion for constructing models in larger systems under extremely non-equilibrium conditions.    
%The values of $F^{(3)}$ decomposed into atomic layers are shown in \cite{Suppl}.
%Layer dependency of $F^{(3)}$ is more complicated even in the case of BOMD.
Strongly enhanced suppression of harmonic terms on surface layers in EMD indicates the emergence of excited electronic states missing from the BOMD framework and we discuss this point next.  
Hereinafter, we restrict our discussion to the non-uniform force reduction of $F^{(2)}_{zz}$.
%\subsubsection{The Hellmann-Feynman force analysis -- Screening by HHO-plasmon complex}
%大事なことは何かというと、こうした電子状態への解析から、より厚いスラブ構造へ今回の成果を外挿できるようになること

%We have hitherto analyzed lattice properties of our system.
%It is indispensable to investigate the physical origins of spatially non-uniform behavior to extract mechanisms general enough to be extrapolated to larger systems.
%Here we focus on the high frequency component of the Hellmann-Feynman force which we omitted when we fit the model parameters.
%The frequency spectrum of force is contemplated to capture the interaction of ions with electronic excitations oscillating much faster than ionic motions.

\begin{figure}%[htbp]
\begin{tabular}{cc}
	\begin{minipage}{0.5\hsize}
	%\begin{overpic}[grid,width=5cm,angle=90]{./Force_Beta.eps}
	\begin{overpic}[width=5cm,angle=90]{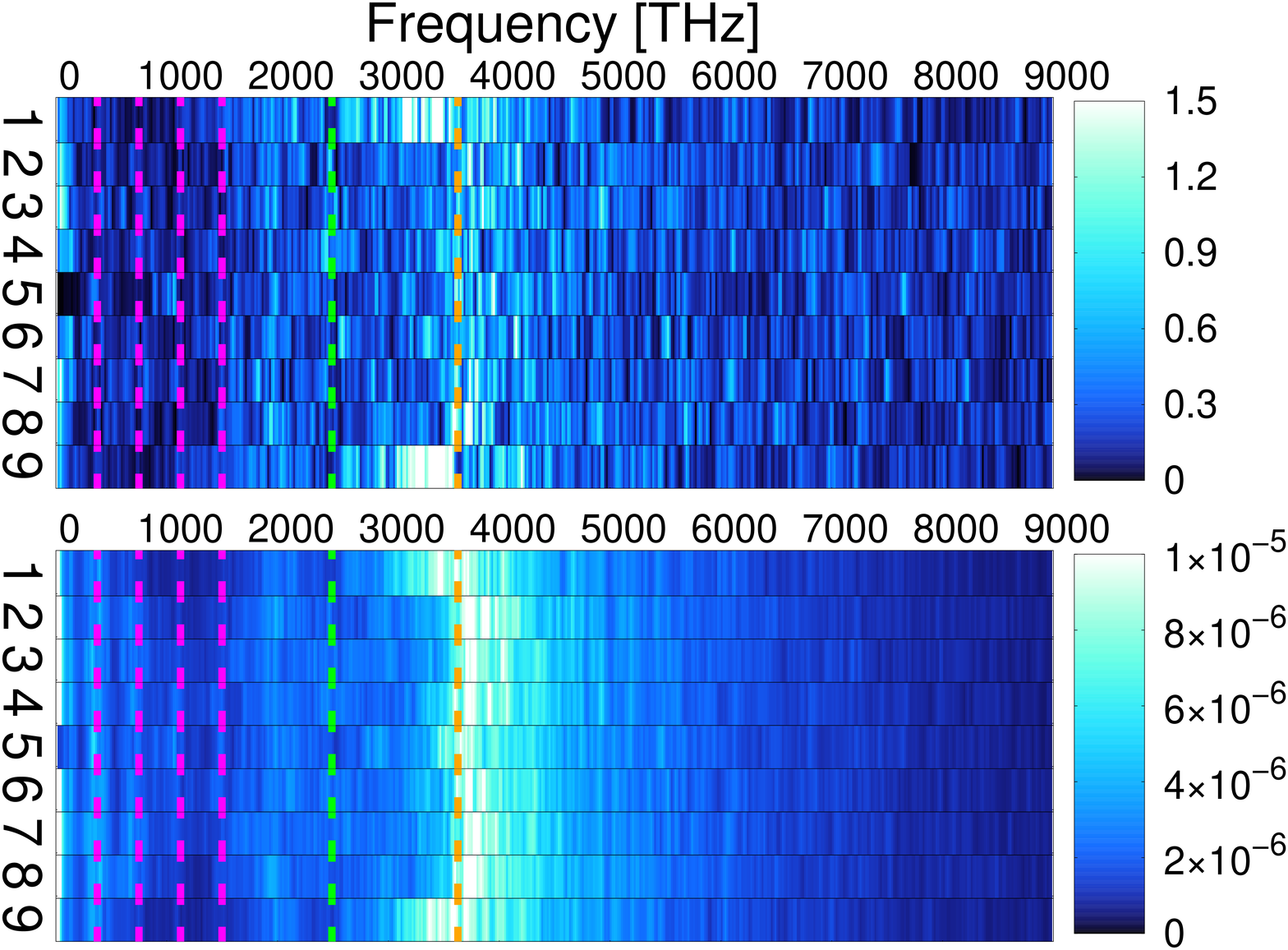}
	\put(-4,82){(a)}
	%\put(-1,30){(c)}
	\end{overpic}
	\end{minipage}
	&
	\begin{minipage}{0.5\hsize}
	%\begin{overpic}[percent,grid,width=4cm]{./Screening.plasma.eps}
	\begin{overpic}[percent,width=4cm]{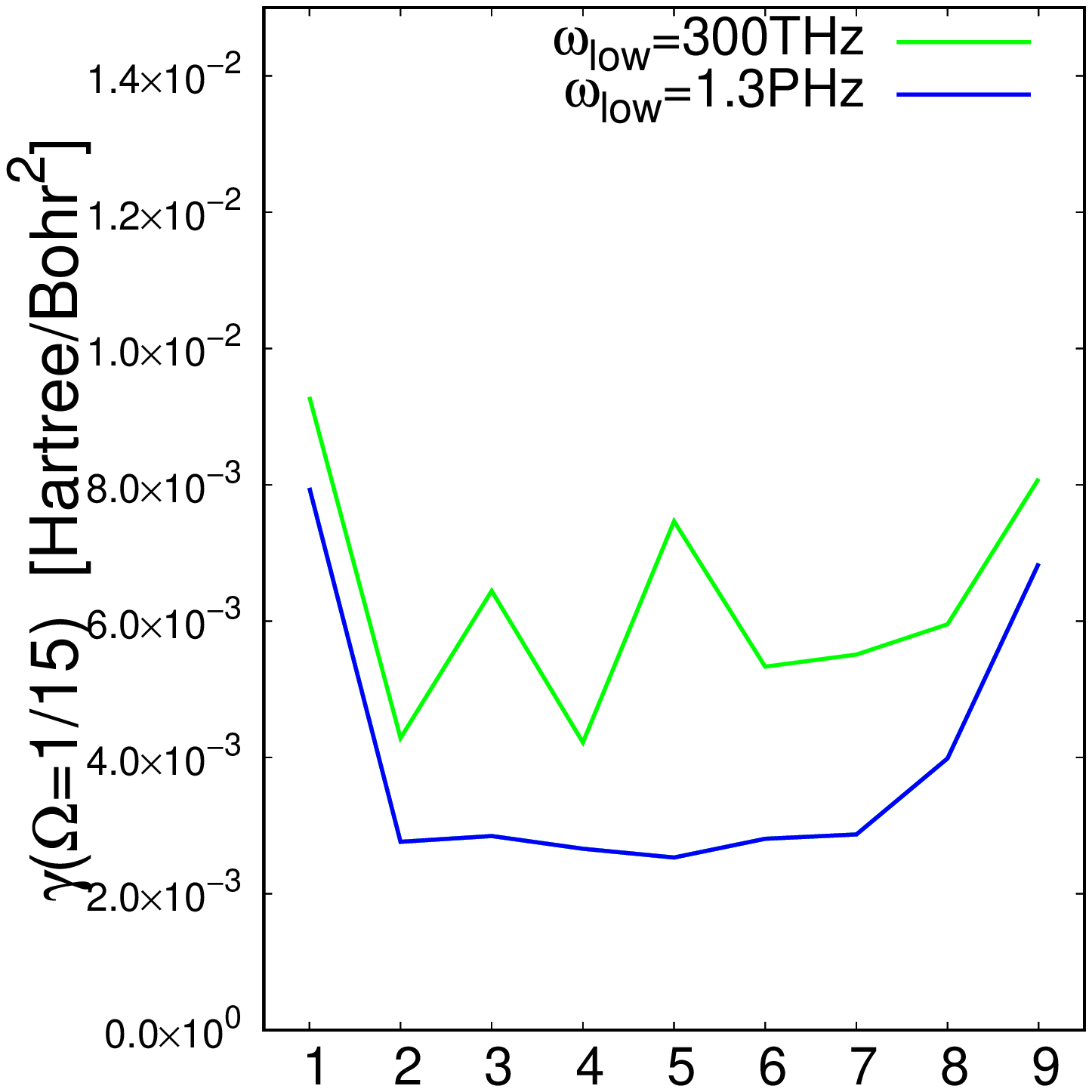}
	\put(0,88){(b)}
	\end{overpic}
	\end{minipage}
\end{tabular}
\caption{(a) Ion-charge density coupling $\Omega\beta_{m}(\omega)$ (left panel) and spectrum of the Hellmann--Feynman force for $E$=3.0 V/\AA~(right panel). 
Integer multiples of the incident laser frequency (magenta dotted lines), $\omega = m\omega_{ph}$ where $\omega_{ph}=375$  TH$_{\rm Z}$, are shown up to $m=4$.
Experimental values of the surface plasmon (green dotted line) and bulk plasmon frequencies (orange dotted line) are shown as a guide to the eye. 
(b) Screening $\gamma_{m}(\omega)$ integrated over frequency $\omega$ from 300  TH$_{\rm Z}$ to 10  PH$_{\rm Z}$ (green line) and from 1.3  PH$_{\rm Z}$ to 10  PH$_{\rm Z}$ (blue line) when $\Omega$=1/15 for a maximum field amplitude of $E$=3.0 V/\AA.
In (a) and (b), the horizontal axes indicate the indices of atoms counted from an outermost atom.}
\label{fig:FFT}
\end{figure}

We show the frequency spectrum of the Hellmann--Feynman force $|f_{\omega,m}|$ for the $m$-th atom along the $z$ direction in the right panel of Fig.~\ref{fig:FFT}(a) for a maximum field amplitude of 3.0 V/$\mathrm{\AA}$.
This spectrum was obtained by averaging 1000 spectra of 60 fs long MD data randomly sampled from the last 360 fs of the 392 fs long MD simulation. 
%These results are obtained by running EMD simulation for 390 fs with 3.63 attosecond time step. 
In the sub-petahertz region, peaks  commensurate with the frequency of the laser pulse $\omega_{ph}$ = 375  TH$_{\rm Z}$ and its integer multiples $m\omega_{ph}$, namely, the high harmonic oscillation (HHO), up to $m=3$ were confirmed.
Furthermore, in the region above 2  PH$_{\rm Z}$, a very large peak was observed.
It is plausible to regard this as plasmonic oscillations relative to the frequency of the volume plasmon and surface plasmon. 

We next examined the screening effect of force constants induced by these high-frequency components. 
It can be easily verified that, when a pair of harmonic oscillators is linearly coupled, the frequency of one is screened while that of the other remains almost unchanged, if the frequency ratio of the two oscillators is very large.   
We generalize this concept and consider a phenomenological interaction $V'$ where the electronic charge density $\rho(\bm{r},t)$ linearly couples to the atomic position $u_{m,z}(t)$ with a coupling constant $\beta_m(\omega)$ as 
\begin{eqnarray}
V' &=&\sum_m \int^{\infty}_0 d\omega \int d\bm{r} \beta(\bm{r},\omega)u_{m,z}\rho_\omega(\bm{r},t)\\
	&\simeq&\sum_m \int^{\infty}_0 d\omega \beta_{m}(\omega)u_{m,z}\rho_\omega(\bm{u}_m,t)\Omega
\label{eq:Vint}
\end{eqnarray}
where $\rho(\bm{r},t) = \int d\omega \rho_{\omega}(\bm{r},t)= \int d\omega \rho_{\omega}(\bm{r})e^{-i\omega t}$ and the index $m$ indicates atoms.
We also used the notation $\beta_m(\omega) = \beta(\bm{u}_m,\omega)$.
In Eq.~(\ref{eq:Vint}), we approximated the space integrals  of $\beta(\bm{r},\omega)\rho_\omega(\bm{r},t)$ by their values at the $m$-th atomic position as $\beta(\bm{u}_m,\omega)\rho_\omega(\bm{u}_m,t)\Omega$, where $\Omega$ is a fitting parameter.
This procedure corresponds to approximating the ion-charge density interaction by a box potential and neglecting the spatial dependency of $\rho_\omega(\bm{u}_m,t)$; thus, $\Omega$ represents the volume of the box potential.
Since the frequencies of the HHO and plasmonic peaks in Fig.~\ref{fig:FFT}(b) are far higher than typical phonon frequencies, the screening effect can be approximated by $\gamma_{m}(\omega)$ in the equation of motion for the $m$-th atom as follows:
\begin{eqnarray}
\ddot{u}_{m,z} &=&-2\sum_nF^{(2)}_{mn}\xi_{mn,z} + \int ^{\omega_2}_{\omega_1}d\omega\gamma_{m}(\omega)u_{m,z}\nonumber \\
&+& O(F^{(3)})
\label{eq:screen}
\end{eqnarray}
where $\gamma_{m}(\omega) = \Omega^2\beta_{m}(\omega)^2/\omega^2$, and $\omega_1$ and $\omega_2$ are frequency cutoffs. 
We abbreviate the third-order terms as $O(F^{(3)})$.
To derive Eq.~(\ref{eq:screen}), we assumed that the harmonic-potential-type restoring force $-\omega^2\rho_\omega$ acts on $\rho_{\omega}$ in its classical equation of motion as $\ddot{\rho}_{\omega}(\bm{r},t) = -\omega^2\rho_\omega(\bm{r},t) - \partial V'/\partial \rho_{\omega}(\bm{r},t)$.
The detailed derivation of $\gamma_{m}(\omega)$ is provided in Sec.~S.III of the Supplemental Material \cite{Suppl}.
We used the value of the charge density integrated over the $xy$ plane of the unit cell to obtain the linear density per bohr at the $m$-th atomic position.
The coupling $\beta_{m}(\omega)$ was computed using $\Omega\beta_{m}(\omega) = <\rho_{\omega}(\bm{u}_m)^*f_{\omega,m}>/<|\rho_{\omega}(\bm{u}_m)|^2>$, where $<\cdots>$ denotes taking the average over randomly sampled spectra as discussed earlier to determine $|f_{\omega,m}|$ in the right panel of Fig.~\ref{fig:FFT}(a) obtained by EMD.
We show $\Omega\beta_{m}(\omega)$ in the left panel of Fig.~\ref{fig:FFT}(a) and $\gamma_{m} = \int d\omega \gamma_{m}(\omega)$ in Fig.~\ref{fig:FFT}(b) for $E=$3.0 V/\AA~by assuming a common $\Omega$ for all atoms.
The values for $\gamma_m$ at lower field amplitude are shown in Fig.~S.2 of Sec.~S.III  of the Supplemental Material \cite{Suppl}. 
We computed $\gamma_{m}$ for $\omega_1$ = 300  TH$_{\rm Z}$ and 1300  TH$_{\rm Z}$, where the latter case omits the contribution of optical frequency $m\omega_{ph}$. 
The screening $\gamma_{m}$ was enhanced on the surface atoms, mostly due to the plasmonic component whose peak positions were lower than those of the inner layers as shown in Fig.~\ref{fig:FFT}(b).
This behavior coincides with the spatial non-uniformity of $F^{(2)}_{zz}$ in Fig.~\ref{fig:force}.
Thus, we conclude that surface-enhanced plasmonic screening of the interatomic force caused the non-uniform spatial dependency.
Although we expect that the plasmonic excitation and HHO also contribute to the behavior of $F^{(3)}$, clarifying these effects will require consideration of the higher-order coupling of $u_{m,z}$ and $\rho_\omega(\bm{r},t)$ in our model. 
This will be investigated in our future work.

Thus far our analysis has clarified the significance of plasmonic effects for modeling the evolution of ablation processes at the sub-picosecond time scale.
By deducing the physical origin of the harmonic force constant reduction to the plasmonic excitations, we can discuss possible finite size effects in larger systems that are too computationally expensive to treat.
Increasing the slab thickness will cause red shift of the plasmonic peaks as it weakens the confinement effect. 
This may enhance the screening effect of interatomic force owing to the $\omega^{-2}$ dependency of $\gamma_m(\omega)$.
Weak confinement will also make the spatial dependency of the $F^{(2)}$ reduction rather moderate. 

%CONCLUSION
Interatomic force constants are one of the most fundamental quantities of lattice systems, upon which the micro- and macroscopic quantities of crystals, such as the dispersion and lifetime of phonons, heat capacity, and diffusion coefficient of energy, rely.
Laser-induced modulation of these quantities is critical to understanding laser ablation processes, and the current work has quantified the modulation of the force constants for both harmonic and anharmonic terms for the first time based on the TDDFT approach.
At this ultrafast timescale and non-equilibrium conditions, collective electronic excitations such as plasmons and HHO take the place of the thermalized electrons that play the main role in ordinary BOMD.
According to our analysis, the non-uniformity of the interatomic force reduction can be ascribed to the non-uniform force screening by plasmons.
This interpretation is consistent with the plasmon-driven mechanism of periodic structure formation at the sub-wavelength scale during ablation processes \cite{RCHP02,BH03,BSRHTRC06,MM08,VMG07,BRK09}.
The investigation of larger systems would be of great interest to us. 
However, at present such studies are hindered by high computational cost, and hence a phenomenological model may need to be developed to describe the force screening effect.
TDDFT is one of the most promising approaches for constructing such models.  

%\subsubsection{Acknowledgement}
This paper is based on the results obtained from the NEDO project ``Development of advanced laser processing with intelligence based on high-brightness and high-efficiency laser technologies'' (TACMI project). The numerical results described in this Letter were obtained using the supercomputing resources at the Cyberscience Center of Tohoku University.
\bibliography{reference}
\end{document}